\documentclass[sigplan]{acmart}

\renewcommand\footnotetextcopyrightpermission[1]{}
\usepackage{listings}
\usepackage{algorithm}
\usepackage{algorithmic}
\usepackage{textcomp}
\usepackage{xcolor}
\usepackage{hyperref}

\usepackage{lipsum}

\newcommand{\intel}{Intel\textsuperscript{\tiny\textregistered}}

\begin{document}
\pagestyle{plain}
\title{Privacy-Preserving Machine Learning in\\ Untrusted Clouds Made Simple}

\author{Dayeol Lee}
\email{dayeol@berkeley.edu}
\affiliation{
 \institution{UC Berkeley}
}
\author{Dmitrii Kuvaiskii}
\email{dmitrii.kuvaiskii@intel.com}
\affiliation{
 \institution{Intel Labs}
}
\author{Anjo Vahldiek-Oberwagner}
\email{anjo.lucas.vahldiek-oberwagner@intel.com	}
\affiliation{
 \institution{Intel Labs}
}
\author{Mona Vij}
\email{mona.vij@intel.com}
\affiliation{
 \institution{Intel Labs}
}

\begin{abstract}
    We present a practical framework to deploy privacy-preserving machine learning (PPML) applications in untrusted clouds based on a trusted execution environment (TEE).
    Specifically, we shield unmodified PyTorch ML applications by running them in \intel\ SGX enclaves with encrypted model parameters and encrypted input data to protect the confidentiality and integrity of these secrets at rest and during runtime. 
    We use the open-source Graphene library OS with transparent file encryption and SGX-based remote attestation to minimize porting effort and seamlessly provide file protection and attestation.
    Our approach is completely transparent to the machine learning application: the developer and the end user do not need to modify the ML application in any way.
\end{abstract}
\maketitle

\section{Introduction}

Machine Learning (ML) is increasingly utilized in many real-world applications.
ML algorithms are first trained on massive amounts of known past data and then deployed to interpret unknown future data, which allows us to forecast weather, classify images, recommend content, and so on.
As machine learning pervades our daily lives, privacy concerns emerge as one of the key issues about this technology. 
Privacy-Preserving Machine Learning (PPML) protects privacy of sensitive user data as well as the trained model---which constitutes the model owner's Intellectual Property (IP)---while performing ML tasks using techniques such as cryptography~\cite{helen}, hardware technologies~\cite{privado, mpmltee}, and differential privacy~\cite{ml-differential}.

In this work, we focus on protecting the confidentiality and integrity of the input data when the computation takes place on an untrusted platform such as a public cloud virtual machine.
We also protect the model for cases where the model owner is concerned about protecting their IP. 
In general, ML workloads have two phases: training and inference.
Both can be viewed as an application that takes inputs and produces an output.
Training applications take a training dataset as input and produce a trained model.
Inference applications take new data and the trained model as inputs and produce the result (the prediction).
The goal of our work is to allow these applications to run in an untrusted environment (like a public cloud), while still ensuring the confidentiality and integrity of sensitive input data and the model. It is essential to protect the confidentiality and integrity in an untrusted environment for PPML applications, since otherwise privacy may be violated by an untrusted 3rd party.

\begin{figure}[t!]
    \centering
    \includegraphics[width=0.9\linewidth]{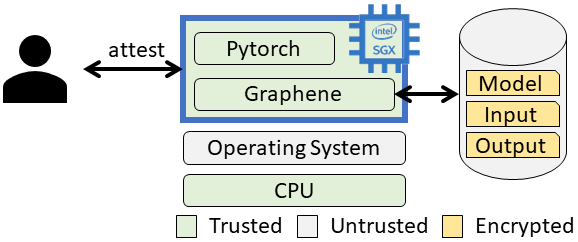}
    \caption{PPML with Graphene and \intel\ SGX.}
    \label{fig:ppml-graphene}
\vspace{-20pt}
\end{figure}

An effective way of achieving this goal is to use \emph{hardware enclaves} such as the ones provided by the \intel\ SGX technology~\cite{sgx}.
SGX enclaves isolate program's execution to protect data confidentiality and integrity, 
and provide a cryptographic proof that the program is correctly initialized and running on legitimate hardware with the latest patches.
By using an SGX enclave, we can perform ML computations on a platform owned by an untrusted party without revealing the actual input data or trained model.
Our approach does not require any changes in the ML application and works in many different scenarios. 

Specifically, we design and implement a system that transparently enforces the confidentiality and integrity of an ML application (see Figure~\ref{fig:ppml-graphene}).
First, we encrypt all confidential inputs with a secret key.
Then, we rely on SGX's remote attestation (via RA-TLS~\cite{ra-tls}) to authenticate to a key server and securely provision a cryptographic key to the SGX enclave.
The unmodified application runs inside the SGX enclave via the Graphene library OS~\cite{graphene-sgx, graphene-website}.
Once the application starts, the inputs are decrypted with the provisioned key, and any output is stored encrypted with the same key.
Since the plaintext inputs and outputs are only accessible inside the enclave, the untrusted platform owner or system software (e.g., operating system) cannot steal them.
To make encryption and decryption transparent to the application, we utilize the Protected File System feature of Graphene.
Using Graphene and its remote attestation with secret provisioning and protected file system techniques, we can build a generic framework to protect the PPML computation seamlessly without changing the actual PPML application.


Note that the developer does not need to change a single line of the Python script; \intel\ SGX and Graphene relieve the developer from this burden by transparently protecting the model, all inputs, and the classification output.

The accompanying tutorial on running the PPML version of PyTorch with Graphene can be found in the official Graphene documentation: \url{https://graphene.readthedocs.io/en/latest/tutorials/pytorch/index.html}. 


\section{PPML Use Cases}

We focus on the scenario where the data owner does not necessarily trust the platform where the computations will take place. As a result, the owner has to protect the confidentiality and integrity of all sensitive data. Here are a few examples of sensitive data:
\begin{itemize}
    \item \textbf{Training with private data}
A streaming company provides a personalized contents recommendation system. The company wants to train its models without collecting user data directly because the customers do not want their data to be used for marketing. In this case, the training ML application may consume user data to train the model but should not allow the data to be used for other purposes.
    \item \textbf{Inference with proprietary model}
A company developed a video editing application with a patented ML-based object detection feature. The application runs on the client side, thus the client's data never leaves the local computer. However, the company is worried that the model could be stolen by competitors.
    \item \textbf{Inference with private data}
A company trained a model that predicts road traffic based on the real-time GPS information of a few nearby users. The company wants to provide a map service using this model. However, collecting real-time location of individuals violates the privacy law. The company needs to make sure that the customers' GPS data will never be disclosed, but still wants to use this data to predict traffic.
\end{itemize}

\section{Background}

The goal of our work is to allow the data owner to selectively encrypt any of the inputs (training data, inference data, proprietary model) and/or the outputs (classification result), and to perform the ML computation on a remote platform while protecting the data from compromised or malicious privileged software. To achieve this goal, we rely on \intel\ SGX ~\cite{sgx} and the Graphene Library OS ~\cite{graphene-sgx, graphene-website}.

\subsection{\intel\ SGX}

\intel\ Software Guard Extensions (SGX) is an ISA extension for \intel\ CPUs that provides capabilities to improve the
confidentiality and integrity of sensitive parts of applications. Whereas the non-sensitive part of an application resides in untrusted memory and executes as usual, sensitive code and data are put inside an SGX enclave---a region of memory that is opaque to all other software including privileged OS/VMM.

The code inside the SGX enclave can execute almost all CPU instructions and can access data both inside and outside of enclave. Attempts by any other software to directly access enclave data from outside of the enclave fail. \intel\ SGX additionally encrypts all enclave data when it leaves the CPU chip so attempts to mount a hardware attack and directly read data in RAM or on memory bus also fail.

Additionally, \intel\ SGX guarantees that the remote execution of a program is cryptographically attested and provides the infrastructure for remote attestation \cite{dcapsgx}. All manufactured SGX processors are listed in the web service maintained by Intel, called Provisioning Certification Service. Cloud providers typically register their SGX platforms with this service and obtain Intel-issued X.509 certificates for these platforms. These certificates are used to prove that the SGX processor is genuine and up-to-date.

\subsection{Graphene Library OS}

Graphene is an open-source library OS that allows an unmodified ELF binary to run on different platforms, including \intel\ SGX enclaves. It serves as a minimal bootloader and an emulation layer between the enclavized application and the underlying host OS. Graphene can be thought of as a minimal re-implementation of the Linux kernel, striving to resolve most application requests (in the form of system calls and opened pseudo-files) inside the SGX enclave and occasionally resorting to host-OS provided resources (network IO, file system IO, thread scheduling, process creation, etc.). Graphene additionally implements extensive checks on system call arguments and return values, helping to protect against maliciously crafted values.

\subsection{Graphene Manifest}

\begin{figure}[t]
    \centering
    \includegraphics[width=0.75\columnwidth]{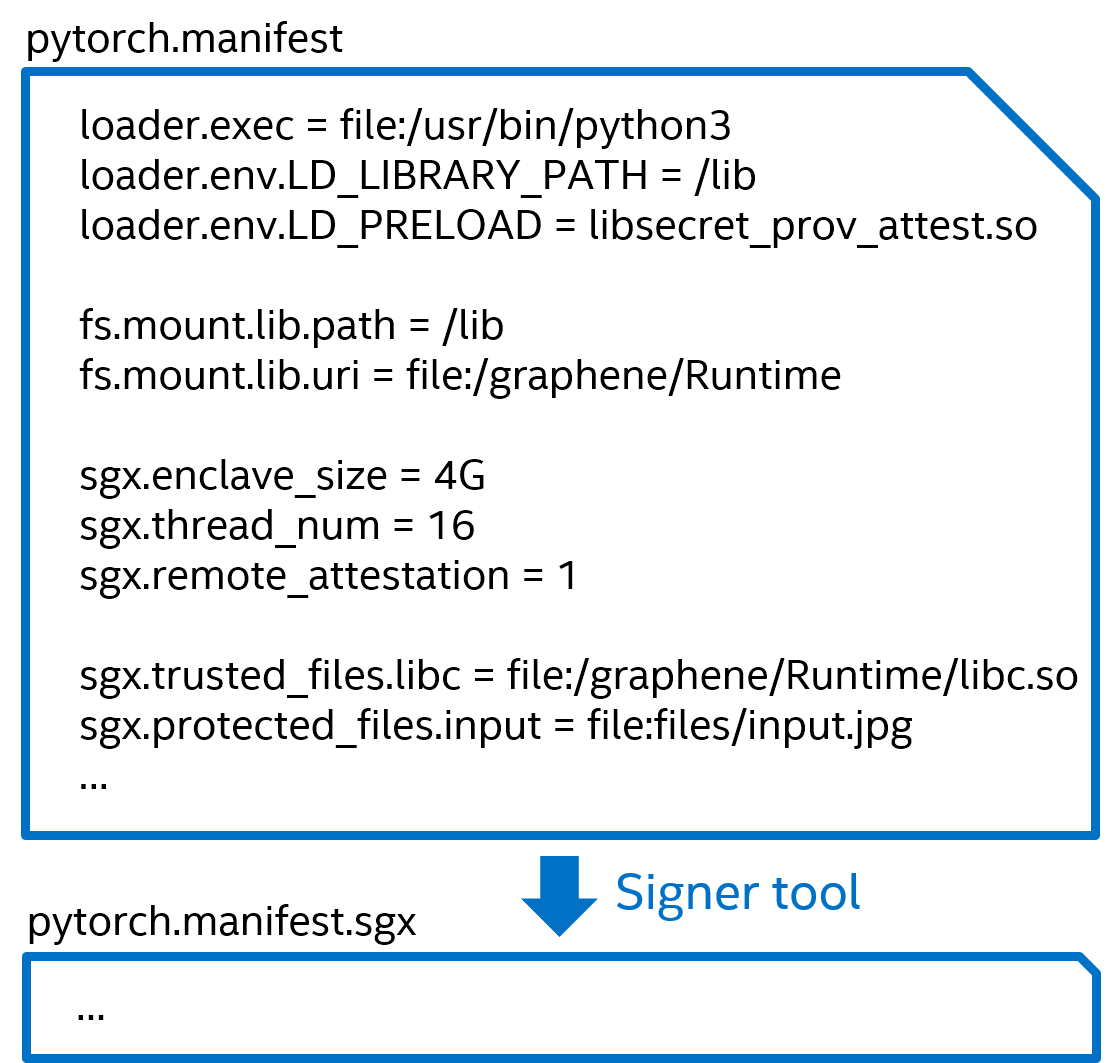} 
    \caption{Graphene manifest file for the enclavized PyTorch application (only a snippet for readability).}
    \label{fig:manifest}
\end{figure}

The application running inside Graphene must be accompanied by a security manifest (see Figure \ref{fig:manifest}). The manifest is a simple configuration file that describes the execution environment of the enclavized application including its security posture, environment variables, loaded libraries, arguments, etc. For example, the manifest specifies the executable to load and run – in our PyTorch PPML case, the Python3 executable. The manifest also specifies a subset of host-OS directories to be mounted inside the enclave so that only they are visible to the application (possibly under a different name). The manifest also contains SGX-specific variables like the maximum enclave size and the maximum number of enclave threads. Finally, files to be consumed by the SGX enclave must be marked as trusted or protected, such that Graphene can verify that these files were not modified during run time.

To verify the manifest, properly finalize it and add the SGX-specific information, Graphene provides the \emph{Signer} tool that automatically generates the final manifest from the user-provided template. This manifest must be shipped together with the application binaries to the remote SGX platform in order to start the SGX enclave with Graphene and the application.


\subsection{Secret Provisioning in Graphene}

Before the remote user provisions her application to the untrusted platform, she must gain trust in the \intel\ SGX hardware available on this platform and in the SGX enclave that runs the PPML application. This is achieved via the Remote Attestation (RA) infrastructure.  

Graphene provides two libraries that transparently add RA flows to the user application. (1) The \emph{RA-TLS library} \cite{ra-tls} augments normal SSL/TLS sessions with the SGX-specific X.509 certificate containing all required measurement information on the attested platform and the SGX enclave and installs an SGX-specific handshake callback to trigger verification of these measurements at the remote user side. (2) The \emph{Secret Provisioning library} builds on top of RA-TLS and automatically establishes a secure SSL/TLS session between the SGX enclave and the remote user so that the user may gain trust in the remote enclave and provision secrets to it before the application starts.

Typical secrets that the remote user may want to provision to the SGX enclave include: encryption keys (to encrypt/decrypt files, network connections, etc.), user credentials (usernames, passwords, tokens), command-line arguments and environment variables, etc.

\subsection{Protected Files in Graphene}

Graphene implements the Protected File System (Protected FS) to transparently protect application files. Integrity- or confidentiality-sensitive files (or whole directories) accessed by the application must be marked as protected files in the Graphene manifest. New files created in a protected directory are automatically treated as protected. The encryption format used in Protected FS is borrowed from the similar feature of \intel\ SGX SDK \cite{pfs}.

Internally, Protected FS divides the protected file in 4KB data blocks and bundles them together in a Merkle tree. Some of the blocks are devoted to Protected FS metadata such as secure hashes over child blocks, derived keys, integrity-protected filenames, etc. AES-GCM is used as the underlying crypto algorithm. As a performance optimization, Protected FS keeps frequently accessed blocks in an LRU cache.


\begin{figure*}[t!]
    \centering
    \includegraphics[width=1.0\textwidth]{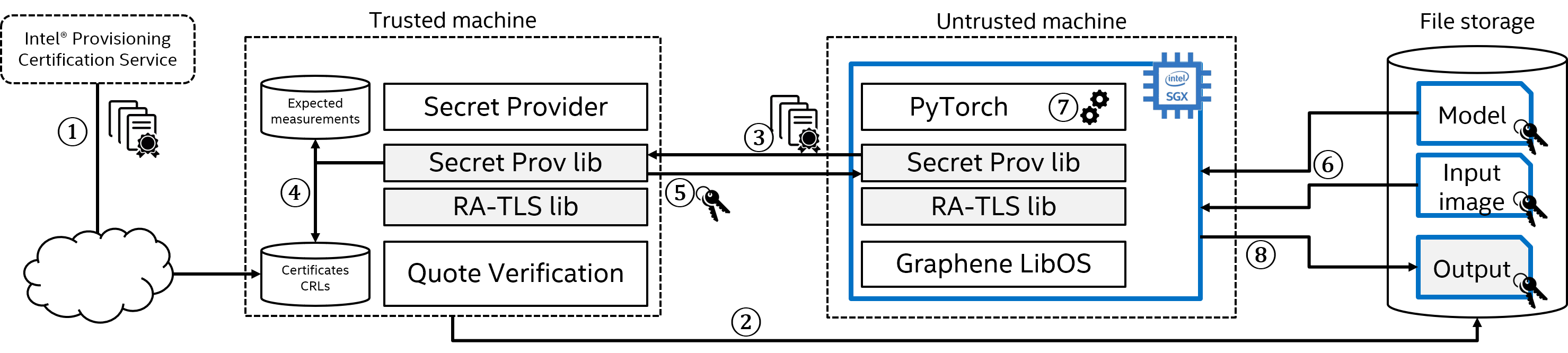}
    \caption{Complete flow to start PyTorch PPML application on an untrusted platform, remotely attest the platform, provision the encryption key for protected files, execute the application, and generate the protected output.}
    \label{fig:workflow}
\vspace{20pt}
\end{figure*}

\section{PPML PyTorch Example}

This section first describes the threat model and then demonstrates the PPML workflow shown in Figure~\ref{fig:workflow}.

\subsection{Threat Model}

We consider the implementation of \intel\ SGX, Graphene Library OS, and the ML application to be trusted. On the other hand, the cloud provider and its system software such as the operating system or virtual machine monitor are untrusted. 

The key manager/secret provider runs on a trusted machine on-premise of the user. 
Its operation is trusted to release cryptographic keys only after the remote attestation successfully proved the identity of the remote SGX enclave. 
An adversary cannot access the trusted machine directly or via vulnerabilities in the secret provisioning application.

Instead, an adversary has control over the untrusted cloud machine. By design of \intel\ SGX, the adversary cannot access the memory contents of any enclave. In addition, any confidential data (such as the secret inputs or ML model) is stored persistently only in encrypted form, preventing the attacker from accessing said secrets.

We assume known attacks---including side-channel attacks---against \intel\ SGX \cite{foreshadow,sgxstep} to be mitigated in hardware.

\subsection{PPML Workflow}

For our PPML use case, we use the PyTorch example deployed with Graphene and running inside an \intel\ SGX enclave. We use the preloaded Secret Provisioning library to receive the cryptographic key for encrypted input and model files. These files are stored on the untrusted cloud storage and are transparently decrypted by Protected FS.

Figure \ref{fig:workflow} shows the complete PPML workflow. To run her application on a particular SGX platform, the remote user must first retrieve the corresponding SGX certificate from the \intel\ Provisioning Certification Service, along with Certificate Revokation Lists (CRLs) and other SGX-identifying information \textcircled{1}.

As a second preliminary step, the user must encrypt the input and model files with her cryptographic key and send these protected files to the cloud storage \textcircled{2}.

Next, the remote platform starts the PPML application inside of the SGX enclave. Meanwhile, the user starts the secret provisioning application on her own machine. The two machines establish a TLS connection using RA-TLS \textcircled{3}, the user verifies that the remote platform has a genuine up-to-date SGX processor and that the application runs in a genuine SGX enclave \textcircled{4}, and finally provisions the cryptographic key to this remote platform \textcircled{5}. Note that during build time, Graphene informs the user of the expected measurements of the SGX application.

After the cryptographic key is provisioned, the remote platform may start executing the application. Graphene uses Protected FS to transparently decrypt the input and the model files using the provisioned key when the PPML application starts \textcircled{6}. The application then proceeds with execution on plaintext files \textcircled{7}. When the PPML application is finished, the output file is encrypted with the same cryptographic key and saved to the cloud provider's file storage \textcircled{8}. At this point, the protected output may be forwarded to the remote user who will decrypt it and analyze its contents.
\section{Conclusion}

In this paper, we presented a framework for building practical privacy-preserving machine learning (PPML) applications and deploying them in an untrusted cloud. Our framework relies on the \intel\ SGX hardware technology and the Graphene LibOS software. We highlighted the ease of use of the framework based on an example PyTorch ML application.
We believe that our framework may be used as a basis for secure PPML and allows to achieve confidentiality, integrity, and privacy of user data (conditioned upon the careful implementation of ML algorithms to preserve privacy) with little porting and deployment effort.

The accompanying tutorial on running the PPML version of PyTorch with Graphene can be found in the official Graphene documentation: \url{https://graphene.readthedocs.io/en/latest/tutorials/pytorch/index.html}. 

\bibliographystyle{ACM-Reference-Format}
\bibliography{main}

\end{document}